\documentclass[12pt,preprint]{aastex}
\shorttitle{Gl 581b}
\shortauthors{L\'opez-Morales et al.}

\begin{document}

\title{Limits to Transits of the Neptune-mass planet orbiting Gl 581\altaffilmark{1}}

\author{Mercedes L\'opez-Morales\altaffilmark{2,3,4}, Nidia I. Morrell\altaffilmark{5}, R. Paul Butler\altaffilmark{2}, Sara Seager\altaffilmark{2}}

\email{mercedes@dtm.ciw.edu, nmorrell@lco.cl, paul@dtm.ciw.edu, seager@dtm.ciw.edu}

\altaffiltext{1}{Based on observations obtained with the 1 m Swope telescope at Las Campanas Observatory, which is operated by the Carnegie Institution of Washington}
\altaffiltext{2}{Carnegie Institution of Washington, Department of Terrestrial
Magnetism, 5241 Broad Branch Rd. NW, Washington D.C., 20015, USA}
\altaffiltext{3}{Carnegie Fellow}
\altaffiltext{4}{American Philosophical Society's Lewis \& Clark Field Scholar in Astrobiology}
\altaffiltext{5}{Las Campanas Observatory, Observatories of the Carnegie Institution of Washington, Casilla 601, La Serena, Chile}

\begin{abstract}

We have monitored the Neptune-mass exoplanet-hosting
M-dwarf Gl 581 with the 1m Swope Telescope at Las Campanas Observatory
over two predicted transit epochs. A neutral density filter centered
at 550nm was used during the first epoch, yielding 6.33 hours of
continuous light curve coverage with an average photometric precision
of 1.6 mmags and a cadence of 2.85 min. The second epoch was
monitored in B-band over 5.85 hours, with an average photometric
precision of 1.2 mmags and 4.28 min cadence. No transits are
apparent on either night, indicating that the orbital inclination is
less than 88.1 deg for all planets with radius larger than 0.38
$R_{Nep}$ = 1.48 $R_{Earth}$. Because planets of most reasonable
interior composition have radii larger than 1.55 $R_{Earth}$ we place
an inclination limit for the system of 88.1 deg. The corresponding
minimum mass of Gl 581b remains 0.97 $M_{Nep}$ = 16.6 $M_{Earth}$.
\end{abstract}


\section{Introduction} \label{sec:intro}
Bonfils et al. (2005a) recently reported the detection of a
Neptune-mass planet around the M-dwarf Gl 581. This planet, with an
orbital period of 5.366 days and minimum mass $M_{2}sini$=
0.97$M_{Nep}$, is only the fifth one found around M-dwarfs. The two
other M-dwarfs currently known to host planets are Gl 876, with two
Jupiter-mass (Delfosse et al. 1998; Marcy et al. 1998) and one
Neptune-mass (Rivera et al. 2005) planets, and Gl436 with a
Neptune-mass planet (Butler et al. 2004). Therefore, GJ581b is only
the third Neptune-mass class planet found around an M-dwarf.

Of the three Neptune-mass class planets, GJ876d and GJ436b have been
photometrically searched for possible transits (Rivera et al. 2005;
Butler et al. 2004), although no transits have been found. GJ581b, on
the other hand, has not yet been searched for transits. The
probability of transit for GJ581b is only 3.3\%, assuming a stellar
radius of 0.29$R_{\sun}$ and an orbital separation of 0.041 AU
(Bonfils et al. 2005a). However, if a transit does occur, the planet
will cover a larger fraction of the M-dwarf than in the equivalent
case of a planet orbiting a Sun-like star. Therefore, transits of GJ
581b would be in principle deeper and easier to detect. The detection
of a transit would allow us to establish the absolute mass of the planet
and its radius. From these parameters we could also determine the mean
density of the planet and also estimate its chemical composition.

GJ 581 (V* HO Lib), an M3V star, has been catalogued as a BY Draconis
variable. This type of variable is characterized by quasiperiodic
photometric variations over time scales from less than a day to
months, and amplitudes ranging from a few hundredths of a magnitude to
0.5 mags. This variability is generally attributed to surface spots
and chromospheric activity, phenomena very common among M-type
dwarfs. Weis (1994) monitored GJ 581 over eight years and found
seasonal and long-term photometric variations of 6 -- 8 mmags in V, R, and I band. However, no short-term (in the
scale of hours) variability measurements have been reported so
far. Furthermore, no information is available about the photometric
precision of the star at wavelengths bluer than V.

This paper presents the first short-term precision light curves of GJ581 over two predicted planetary transit epochs. The light curves were collected in optical and B bands and span over 6.33 and 5.85 consecutive hours, respectively. Section 2 describes the observations, with their analysis presented in section 3. The results of a geometrical search for transits are presented in section 4.

\section{Observations} \label{sec:obs}

We measured the light curves between 2:11 and 8:31 Apr 24 2006 UT and 2:22 and 8:44 May 10 2006 UT at the Henrietta Swope telescope, located at Las Campanas Observatory in Chile. Transits were predicted to occur around 2:54 Apr 24 UT and 5:16 May 10 UT (mid-transit times), based on the ephemeris given by Bonfils et al. (2005a). The estimated minimum duration of the transits is $\sim$ 85 min. The uncertainty on the estimated mid-transit times is $\sim$ 86.4 min.

The Swope is currently equipped with a 2048 x 3150, 15$\micron$ pixel SITe CCD that provides a field of view of 14'.8 x 22'.8. The dynamic range of the CCD is 32,727 ADU (analog-to-digital converter units), with a gain of 2.5 $e^{-}$ per ADU. We used a 2048 x 2048 (14'.8 x 14'.8) subraster of the CCD to reduce the readout time to 128s (1 x 1 binning) and therefore increase the duty cycle of our observations. The aperture stop described in L\'opez-Morales (2006) to avoid saturation of bright stars has been now replaced by a 10$\%$ transmission neutral density filter (ND0.9) with a 3000$\AA$ FWHM passband centered at 5500$\AA$. We have also added to the setup a standard Johnson B-band filter.

GJ~581 (V = 10.56; B-V = 1.61) was strategically placed on the CCD to include a suitable comparison star in the frames. As comparison we used the nearby star BD-06 4172 (V = 10.50; B-V = 1.30), located at a distance of $\Delta\alpha$ = $0^{m}$.896 and $\Delta\delta$ = 12'.747 from GJ~581. As a check star we used a slighty fainter object (V = 12.2) located at $\alpha(2000)$ = 15:19:27.5, $\delta(2000)$ = -07:31:44.

All the images from the first night were taken through the neutral density filter ND0.9. We collected a total of 133 images with 30s exposure times, covering a range of airmasses between 1.08--1.82. GJ 581 was monitored over 6.33 hours with an average time of cadence of 2.85 min. The images on the second night were collected using the B-band filter. A total of 82 images were collected in this case, with exposure times of 120s and airmasses between 1.08--1.62. This time we monitored the target over 5.85 hours with an average time of cadence of 4.28 min. The photometric precision per frame is $\sim$ 3 $10^{-4}$ mmags, from Poisson noise alone. Equation (10) from Dravins et al (1998) gives a scintillation level of 0.74 mmags in the ND0.9 band and 0.37 mmags in the B-band, for our exposure times, at an intermediate airmass of 1.4 (we used for these calculations a telescope aperture diameter of 100 cm, an observatory height of 2100 m, and an atmospheric scale height of 8000 m). Our photometric precision is therefore limited by scintillation.

\section{Analysis}\label{sec:anal}

Each image was bias substracted and flatfielded using the same combined bias and the combined flats in each filter. Next we performed aperture photometry on the target, comparison, and check stars in each calibrated frame over a series of apertures ranging between 10 and 30 pixels for the ND0.9 data and between 10 and 26 pixels for the B-band data. The area used to compute the sky background around each star was the same in all frames, i.e. a 15-pixel annulus at a radius of 40 pixels from the center of the stars. The sky background around each star was computed using a $\sigma$-clipping algorithm in order to avoid contamination by nearby objects or residual bad pixels. The optimal combination of apertures for each light curve was derived following a procedure analogous to the one described in L\'opez-Morales (2006). 

The average dispersion of the comparison and check stars is of the order of 0.8 mmags in both filters. Small differential extinction effects are apparent in the differential light curve of the target and the comparison. Those effects have been corrected by applying second order extinction correction equations of the form
\begin{equation}
\Delta ND = \Delta nd - k_{nd}\Delta X + c_{nd}\Delta (B-V)
\end{equation}
and
\begin{equation}
\Delta B = \Delta b - k_{b}\Delta X + c_{b}\Delta(B-V),
\end{equation}
\\
\noindent where $\Delta$ indicates the difference in the associated quantities for each star, as derived by Hardie (1962). $\Delta$nd and $\Delta$b are the differences in instrumental magnitude of the target and comparison stars in ND0.9 and B-band, respectively. $\Delta$X is the difference in airmass between the two objects. This value changes with time, depending on the position of the objects in the sky. In our particular case, the absolute value of $\Delta$X varies between 1.5 $10^{-4}$ and 1.3 $10^{-2}$ for the range of airmass covered by our observations. $\Delta$(B-V) is the difference in color between the two stars, in this case $\Delta$(B-V) = 0.3. Finally, $k_{nd}$ and $c_{nd}$, and  $k_{b}$ and $c_{b}$ denote the extinction and color coefficients in ND0.9 and B. The values of those coefficients have been adopted from recent measurements by Hamuy et al. (2006), where they compute average values of extinction and color coefficients at the Swope telescope over $\sim$ 230 nights, in different filter bands. For the ND0.9 filter we used the values of the coefficients in V-band. 

The final light curves are presented in Figure 1. Figure 1a shows the 6.33 hour time series coverage of GJ 581 with the ND0.9 filter, with an average standard deviation of the light curve of 1.65 mmags. The light curve in B-band is represented in Figure 1b. The time coverage of this second light curve is 5.85 hours. The average standard deviation of the light curve in this case is 1.17 mmags. A file containing the data in this Figure is available online. A sample of the contents of that file are illustrated in Table 1. Notice that the observed photometric dispersion in the light curves is larger than expected from the scintillation levels computed in section 2. This discrepancy can be attributed to uncertainties in the estimation of the scintillation noise, low-level systematics in our data, or intrinsic variability of the stars.

\section{Search for Transits} \label{sec:tran}

To determine whether or not a planet transit is present in our data,
we need to estimate the transit duration and transit depth of GJ 581b.  In
turn we need to estimate the radius of the planet, the radius of the
star, the orbital separation between the two objects, and the orbital
period of the system. 

The orbital separation and orbital period have been determined by Bonfils
et al. (2005a), who find GJ 581b to be on a circular orbit around the
star, at a distance of 0.041 AU, and with an orbital period of 5.366
days. Gl 581 is an M3 dwarf with an estimated mass of 0.31 $\pm$ 0.02
$M_{\sun}$ (estimated from stellar mass-luminosity relations; see
Bonfils et al. 2005a). That mass corresponds to a radius of about 0.29
$\pm$ 0.03 $R_{\sun}$, based on the mass-radius relation models by
Baraffe et al. (1998).

The radius of the planet will depend heavily on its interior composition. 
A Neptune-mass planet can
plausibly range from 1.55 to 9.5 $R_{\oplus}$, depending on whether
the object is fully composed of iron (Seager et al., in preparation)
or if it is an evaporating H/He gas planet (Baraffe et al. 2005). The
radius of a planet composed of a large fraction of H/He as close to
the host star as GJ 581b will also be affected by the amount of stellar
luminosity incident on the planet.  This in turn depends on distance
to the star as well as the planet's albedo and energy transportation
mechanism. These last two parameters are currently unknown for
extrasolar planets. At present only an estimation of the albedo of
HD209458b has been reported by Rowe et al. (2006), who find a
3$\sigma$ upper limit for this parameter of $\sim$ 0.25. For all of
the above reasons it is difficult to estimate a radius for GJ 581b a
priori of any measured transits.

Based only on geometry, we can estimate upper limits for the transit 
depth and hence planet radius. Assuming an edge-on configuration 
($i$ = 90 deg) and a
circular orbit, the transit duration can last between 85 to 99 minutes
for the range of planetary radii 1.48--7 $R_{\oplus}$. That range of 
radii also gives transit depths between $\sim$ 2.5  and
54 mmags, assuming full
transits and a uniform luminosity of the surface of the star (i.e., no
limb darkening or star spots).

Figure 2 shows a schematic of full transits of a 1.48 $R_{\oplus}$
(solid line), a 3.9 $R_{\oplus}$ (dashed line), and a 7.0 $R_{\oplus}$
(dotted line) superimposed on our ND0.9 filter and B-band light
curves. These schematic box models do not include any
physical effects associated with either the planet or the star. The
horizontal error bars below each transit illustrate the uncertainty in
the predicted mid-transit time. The 2$\sigma$ limit in the photometric
dispersion of the light curves around the expected time of transit (binned
in a 85 minute time bin) is $\sim$ 2 mmags in both light curves.

It is clear that transits of the illustrated depths and durations do
not occur during either one of the predicted transit windows. From the 2$\sigma$ limit above,  combined with the radius for Gl 876 A, we can discard the presence of full transits of planets as small as 1.48 $R_{\oplus}$ at inclinations higher than 88.1 deg.

Our observations therefore rule out transits of GJ581b for inclinations larger
than 88.1 deg and a planetary radius larger than 1.48
$R_{\oplus}$. Transits could still be present if $R_p < 1.48 R_{\oplus}$
because they are not detectable within the photometric dispersion of
our light curves. Only higher precision, either ground- or
space-based, observations will be able to find out if transits of
planets smaller than our radius detection limit occur. Intrinsic
low-amplitude photometric variability of the star will most likely
impose the final limit to detectable transit depths. We do, however,
note that the star appears to be more photometrically stable at short
time scales than at the time scale of months to years measured by Weis
(1994).

We emphasize that planets of all plausible interior compositions are
larger than 1.5 $R_{\oplus}$. This comes from the mass-radius relation
for cold homogeneous iron planets (Seager et al., in
preparation). Iron is the most dense abundant material from which
planets are formed. Its unlikely that a planet is solely composed of
iron, but any other material will make the planet less dense and hence
larger. Similarly, including temperature affects for a planet in an
0.041~AU orbit around an M star will make the planet larger,
marginally so for a rocky planet and moreso for a gas planet. For
reference, the solar system terrestrial planets have iron cores;
Mercury has a a 60-70 percent iron core by mass and Earth has a 32.5\%
iron core by mass (with 10\% of lighter material).

The absence of detectable transits means the radius and hence nature
of GJ 581b remains unknown. Planets larger than our detection limit of
1.48 $R_{\oplus}$ may include: ``rock giant'' silicate planets like
massive Earths, a water world (akin to Neptune without a gas envelope)
or even an evaporating gas planet.

\acknowledgments

We thank the technical personnel at Las Campanas
Observatory for their help with the instrumental parts of this
project. M.~L-M. acknowledges research and travel support from the
Carnegie Institution of Washington through a Carnegie Fellowship and
from the American Philosophical Society through a Lewis \& Clark
Research Fund in Astrobiology Scholarship. This project has been
partially supported by the National Aeronautics and Space
Administration through grant NAG5-12182.

\clearpage

\begin{table}
\centering
\footnotesize
\caption{Sample of the contents of online table containing the data represented in Figure 1}
\label{tab:AbsDim} 
\begin{tabular}{ccc}
\hline\hline
HJD (days) & $\Delta B$ (mag) & $ \sigma_{\Delta B}$ (mag)\\
\hline

2453865.604493 & -0.0025374 & 0.0000005\\
2453865.607757 & 0.0007653 & 0.0000005\\
2453865.611091 & 0.0005909 & 0.0000005\\
2453865.614100 & 0.0007751 & 0.0000005\\
2453865.616947 & -0.0005518 & 0.0000005\\
...            & ...        & ...      \\
\hline\hline
\end{tabular}
\end{table}

\clearpage

\begin{figure}
\plotone{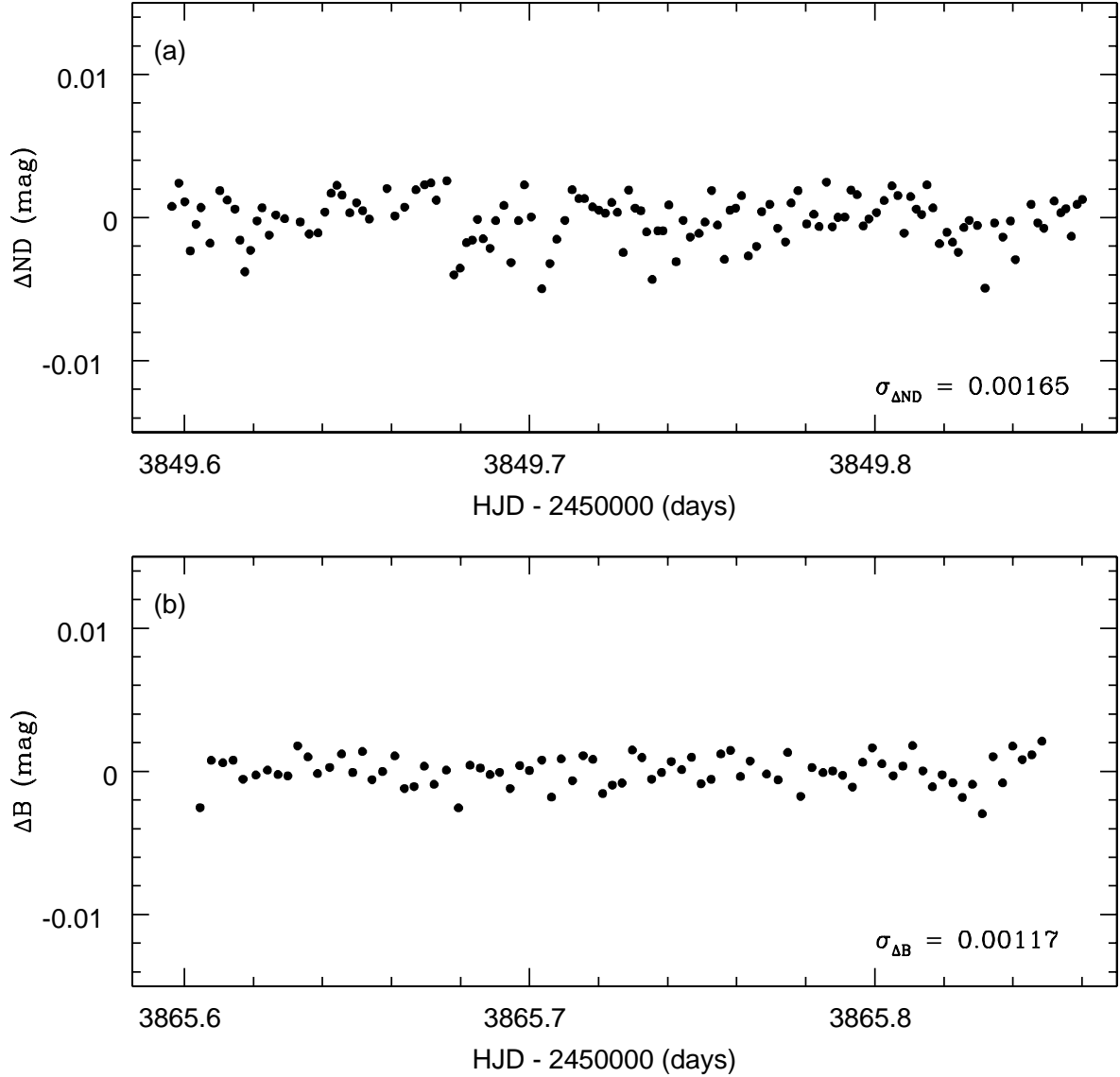}
\caption{ Light curves of Gl 581 with the ND0.9 filter (a) and the B-band filter (b). The time coverage in (a) is 6.33 hours, with an average photometric dispersion of 1.65 mmags. The time coverage in (b) is 5.85 hours, with an average photometric dispersion of 1.17 mmags. }
\label{fig:lcs}
\end{figure} 

\clearpage

\begin{figure}
\plotone{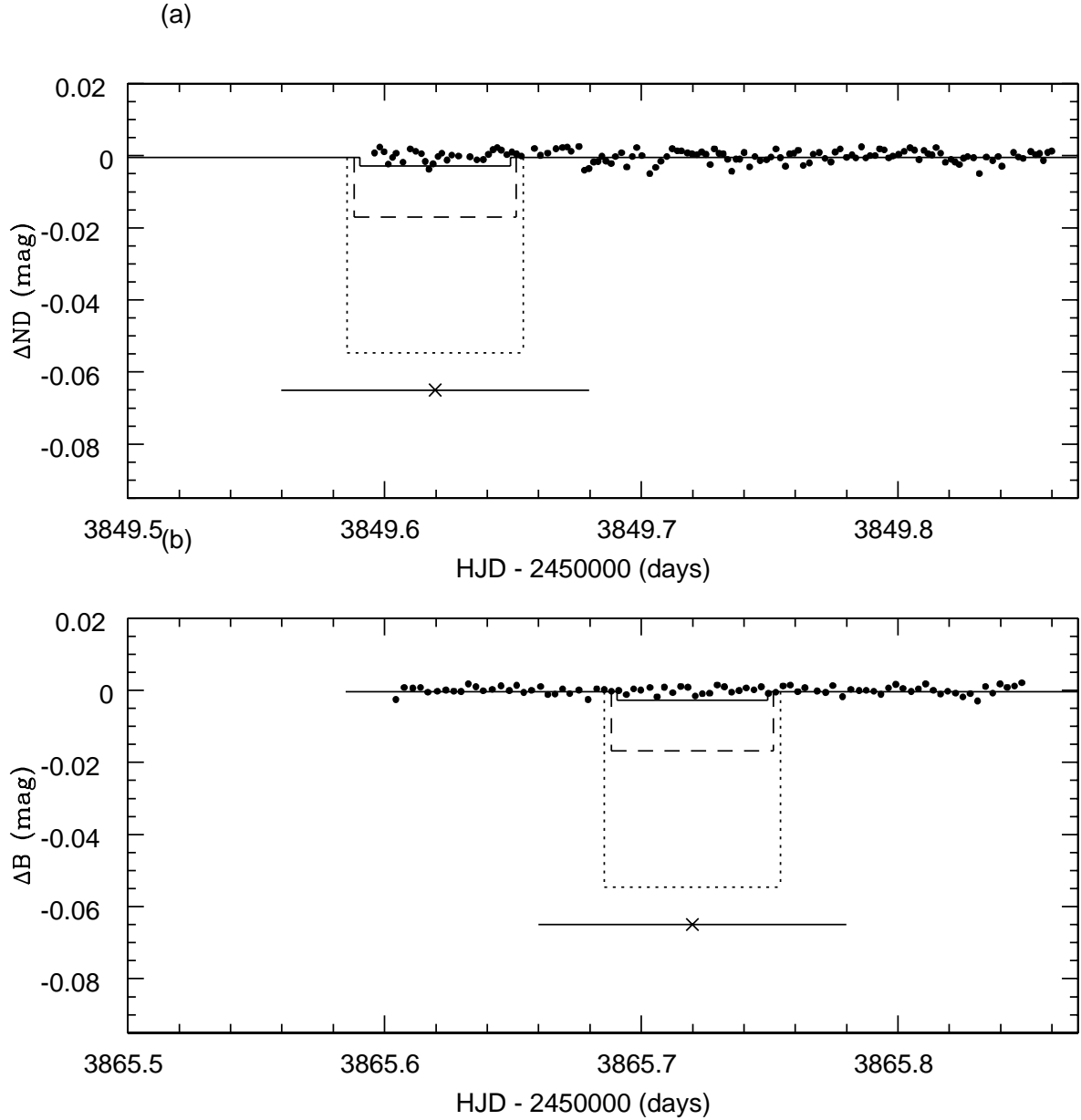}
\caption{ Schematic representation of geometric transits for three possible values of the planet's radius, 1.48 $R_{Earth}$ (solid line), 3.9 $R_{Earth}$ (dashed line), and 7.0 $R_{Earth}$ (dotted line), overplotted on our ND0.9 filter (a) and B-band filter (b) light curves. The solid lines under each transit illustrate the uncertainty in the predicted times of eclipse (86.4 min), with the predicted mid-transit times marked as crosses.}
\label{fig:lcs2}
\end{figure}


\begin{thebibliography}{}

\bibitem[Baraffe et al.(2005)]{2005A&A...436L..47B} Baraffe, I., Chabrier, 
G., Barman, T.~S., Selsis, F., Allard, F., \& Hauschildt, P.~H.\ 2005, 
\aap, 436, L47 

\bibitem[Baraffe et al. (1998)]{bar98} Baraffe, I., Chabrier, G., Allard, F. \& Hauschildt, P. H. 1998, A\&A, 337, 403

\bibitem[Bonfils et al.(2005a)]{bon05a} Bonfils, X., Forveille, T., Delfosse, X., Udry, S., Mayor, M., Perrier, C., Bouchy, F., Pepe, F., Queloz, D. \& Bertaux, J.-L. 2005a, A\&A Letters, 443, 15

\bibitem[Bonfils et al.(2005b)]{bon05b} Bonfils, X., Delfosse, X., Udry, S., Santos, N. C., Forveille, T. \& S'egransan, D. 2005b, A\&A, 442, 635 

\bibitem[Butler et al.(2004)]{but04} Butler, R.P., Vogt, S.S., Marcy, G.W., Fischer, D. A., Wright, J. T., Henry, G.W., Laughlin, G. \& Lissauer, J.J. 2004, ApJ, 617, 580

\bibitem[Delfosse et al.(1998)]{del98} Delfosse, X., Forveille, T., Mayor, M., Perrier, C., Naef, D. \& Queloz, D. 1998, A\&A Letters, 338, 67

\bibitem[Hamuy et al. (2006)]{ham06} Hamuy, M., Folatelli, G., Morrell, N. I., Phillips, M. M., Suntzeff, N. B., Persson, S.E., Roth, M., Gonzalez, S., Krzeminski, W., Contreras, C., Freedman, W. L., Murphy, D. C., Madore, B. F., Wyatt, P., Maza, J., Filippenko, A. V., Li, W. \& Pinto, P. A. 2006, PASP, 118, 839

\bibitem[Hardie (1962)]{har62} Hardie, R.H. 1962, {\it Astronomical Techniques}. Edited by W.A. Hiltner, University of Chicago Press, p180

\bibitem[Dravins et al. (1998)]{dra98} Dravins, D., Lindergren, L., \& Mezey, E. 1998, PASP, 110, 610
\bibitem[L\'opez-Morales(2006)]{lop06} L\'opez-Morales, M. 2006, PASP, 118, 71

\bibitem[Marcy et al.(1998)]{mar98} Marcy, G.W., Butler, R.P., Vogt, S.S., Fischer, D. \& Lissauer, J.J. 1998, ApJ Letters, 505, 147

\bibitem[Rivera et al. (2005)]{riv05} Rivera, E.J., Lissauer, J.J., Butler, R.P., Marcy, G.W., Vogt, S.S., Fischer, D. A., Brown, T.M., Laughlin, G. \& Henry, G.W. 2005, ApJ, 634, 625

\bibitem[Rowe et al. (2006)]{row06} Rowe, J.F., Matthews, J.M., Seager, S., Kuschnig, R., Guenther, D.B., Moffat, A.F.J., Rucinski, S.M., Sasselov, D.D., Walker, G. A. H. \& Weiss, W.W. 2006, astro-ph/0603410  


\bibitem[Seagroves et al.(2003)]{sea03} Seagroves, S., Harker, J., Laughlin, G., Lacy, J. \& Castellano, T. 2003, PASP, 115, 1355

\bibitem[Weis (1994)]{Wei94} Weis, E.W. 1994, AJ, 107, 1135

\end{thebibliography}
\end{document}